\documentclass[twocolumn]{nature}

\usepackage[dvips]{graphicx} 
\usepackage{amsfonts}
\usepackage{amssymb}
\usepackage{amscd}
\usepackage{amsmath}    
\usepackage{enumerate}
\usepackage{epsfig}
\usepackage{subfigure}
\begin{document}

\title{Measurement-device-independent quantum key distribution over untrustful metropolitan network}

\author{Yan-Lin Tang$^{1,2}$, Hua-Lei Yin$^{1,2}$, Qi Zhao$^{3}$, Hui Liu$^{1,2}$, Xiang-Xiang Sun$^{1,2}$, Ming-Qi Huang$^{1,2}$,
Wei-Jun Zhang$^{4}$, Si-Jing Chen$^{4}$, Lu Zhang$^{4}$, Li-Xing You$^{4}$, Zhen Wang$^{4}$,
Yang Liu$^{1,2}$, Chao-Yang Lu$^{1,2}$, Xiao Jiang$^{1,2}$,
Xiongfeng Ma$^{3}$, Qiang Zhang$^{1,2}$, Teng-Yun Chen$^{1,2}$, and Jian-Wei Pan$^{1,2}$}

\maketitle

\begin{affiliations}
\item
Shanghai Branch, National Laboratory for Physical Sciences at Microscale and Department of Modern Physics, University of Science and Technology of China, Shanghai 201315, China

\item
Synergetic Innovation Center of Quantum Information \& Quantum Physics,  University of Science and Technology of China, Hefei, Anhui 230026, China

\item
Center for Quantum Information, Institute for Interdisciplinary Information Sciences, Tsinghua University, Beijing, 100084, China

\item
State Key Laboratory of Functional Materials for Informatics, Shanghai Institute of Microsystem and Information Technology, Chinese Academy of Sciences, Shanghai 200050, China
\end{affiliations}

\maketitle

\begin{abstract}
Quantum cryptography\cite{Bennett:BB84:1984,Ekert:QKD:1991} holds the promise to establish an information-theoretically secure global network\cite{NNews:Leapout:2014}. All field tests of metropolitan-scale quantum networks\cite{elliott:DARPA:2006,Peev:SECOQC:2009,Chen:Metropolitan:2010,Sasaki:TokyoQKD:2011} to date are based on trusted relays. The security critically relies on the accountability of the trusted relays, which will break down if the relay is dishonest or compromised. Here, we construct a measurement-device-independent quantum key distribution (MDIQKD)\cite{Lo:MIQKD:2012} network in a star topology over a 200 square kilometers metropolitan area, which is secure against untrustful relays and against all detection attacks. In the field test, our system continuously runs through one week with a secure key rate ten times larger than previous result\cite{Tang:FieldMDIQKD:2014}. Our results demonstrate that the MDIQKD network, combining the best of both worlds --- security and practicality, constitutes an appealing solution to secure metropolitan communications.
\end{abstract}

\maketitle

\begin{figure*}[tbh]
\centering
\resizebox{12cm}{!}{\includegraphics{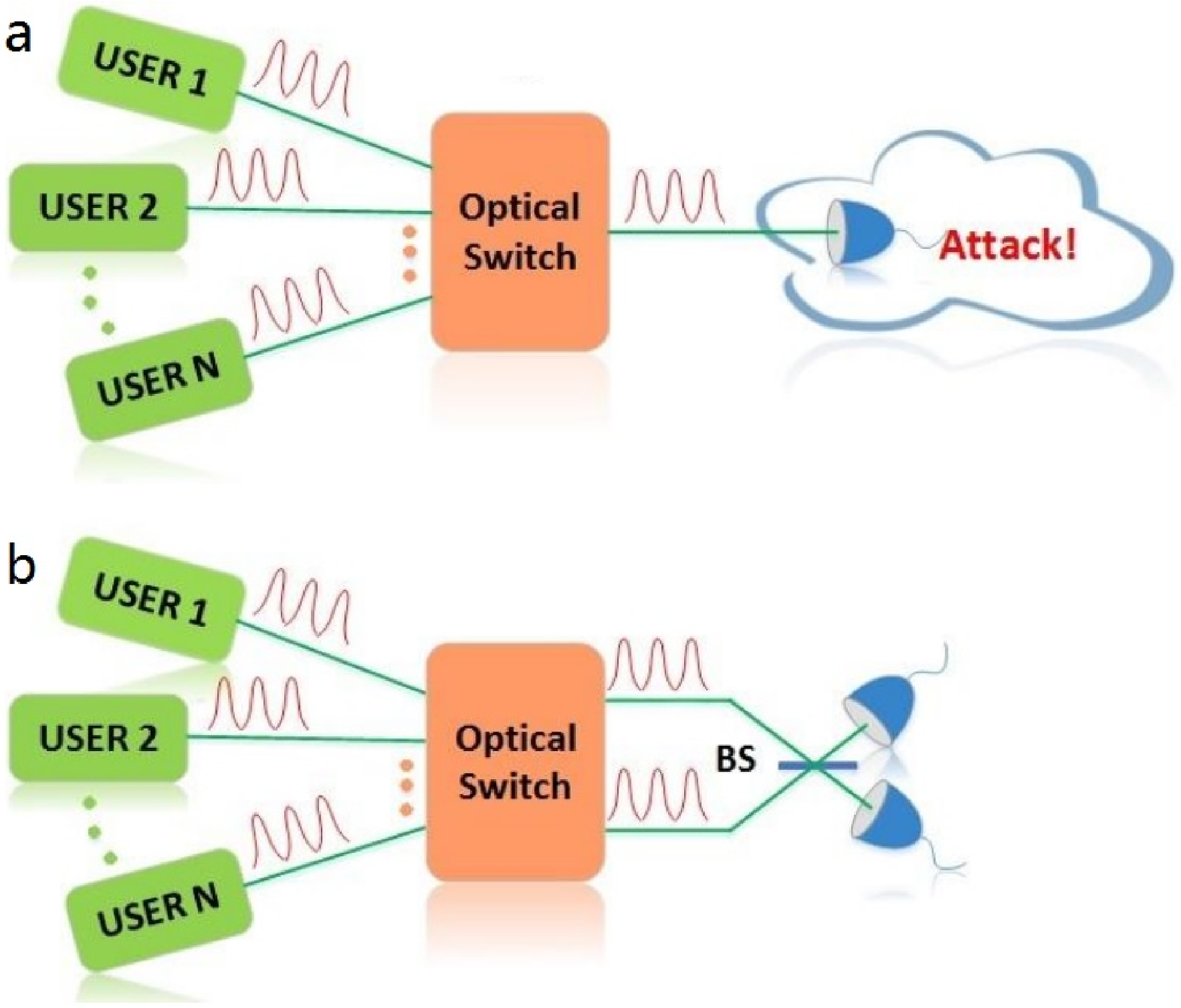}}
\caption{(a) Schematic of conventional QKD network with detectors as the shared resources which are vulnerable to detection attacks.
(b) Schematic of star-type MDIQKD network, in which the shared detectors can be even controlled by Eve, but without any leakage of key information.}
\label{Fig:NetScheme}
\end{figure*}

Quantum key distribution (QKD)\cite{Bennett:BB84:1984,Ekert:QKD:1991} can in principle offer information-theoretical security between two remote parties, guaranteed by the fundamental laws of quantum mechanics. The last three decades have witnessed tremendous advances in both theoretical developments and successful experimental demonstrations of various quantum cryptographic systems. Moreover, QKD networks of various topologies have emerged and extended to more users in larger domains\cite{elliott:DARPA:2006,Peev:SECOQC:2009,Chen:Metropolitan:2010,Sasaki:TokyoQKD:2011,Frohlich:QuantumNetWork:2013}.
To increase the scalability of these networks, the architectures often adopt the idea of sharing, and thus star-type network is preferable for sharing the most expensive resource --- single photon detectors\cite{Frohlich:QuantumNetWork:2013}, as shown in Fig.~\ref{Fig:NetScheme}(a).
With such a topological structure, it is straightforward to directly add more users with low hardware requirement.

From the security point of view, the existing star-type networks have to assume the central relays to be trustful, which is a critical shortcoming. Once the relay is dishonest, the security of the whole network is down. Furthermore, an honest relay with imperfect devices still suffers from various attacks that exploit the loopholes caused by the gap between the idealized devices assumed in the security proof and the realistic ones\cite{GLLP:2004}. Indeed, single photon detectors can be difficult for legitimate users to characterize precisely but easy for technology-advanced eavesdroppers to exploit a certain imperfection to attack\cite{Qi:TimeShift:2007,Lydersen:Hacking:2010}. It is important to note that in all previous network implementations, trustful relays are utilized, which constitutes the weakest point of security.

Remarkably, the MDIQKD protocol\cite{Lo:MIQKD:2012}, inspired by the time-reversed entanglement-based QKD protocol\cite{Biham:1996:Quantum,Inamori:TimeReverseEPR:2002,Stefano:MDIQKD:2012}, can close all the detection loopholes. Besides, MDIQKD is intrinsically suitable for a star-type network architecture with measurement devices placed at the central relay, as depicted in Fig.~\ref{Fig:NetScheme}(b). Since its security doesn't rely on any assumption on measurement, MDIQKD network even allows the eavesdroppers to have a full control of the relay without compromising the security. Therefore, the MDIQKD networks are able to solve the security loophole existing in the conventional star-like QKD networks and almost all existing quantum networks.

Up till now, many efforts have been devoted to proof-of-principle experimental demonstrations of the MDIQKD protocol\cite{Tittel:MDIQKDFielfTest:2013,Liu:MIQKDexp:2013,Silva:DemoPolMDIQKD:2013,tang:experimental:2013,pirandola:CVMDIQKD:2015}. Further experiments, accounting for long distance, high loss and field test, have also been reported\cite{Tang:200kmMDIQKD:2014,valivarthi:60dBMDIQKD:2015,Tang:FieldMDIQKD:2014}.
Nevertheless, these are all limited to point-to-point configurations.
A network implementation, yet to be developed, is crucial for enabling and extending practical applications of MDIQKD.

\begin{figure*}[tbh]
\centering
\resizebox{12cm}{!}{\includegraphics{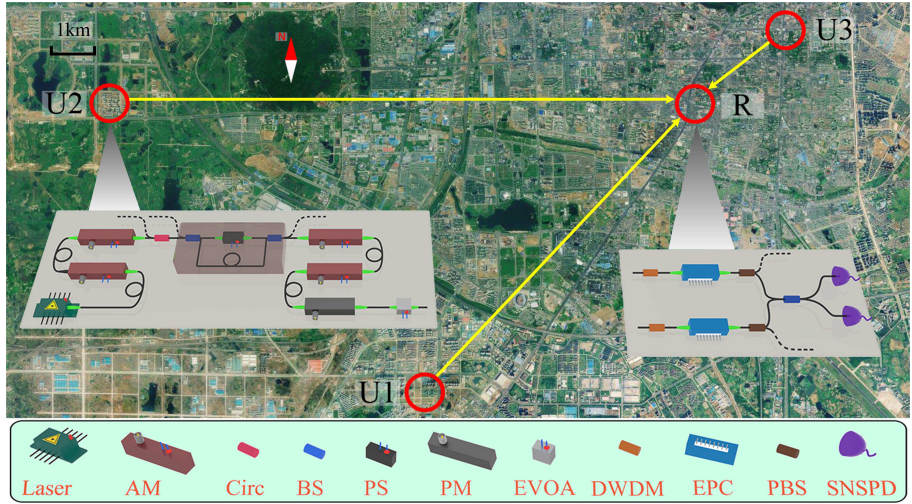}}
\caption{Birds-eye view of the MDIQKD network topology. User $U1$ is placed in an administrative committee (AC) $(N31^{\circ}47^{'}5^{''}, E117^{\circ}12^{'}58^{''})$, user $U2$ in the site of Animation Industry Park in Hefei (AIP) $(N31^{\circ}50^{'}6^{''}, E117^{\circ}7^{'}52^{''})$, and user $U3$ in the site of an office building (OB) $(N31^{\circ}50^{'}57^{''}, E117^{\circ}16^{'}50^{''})$. Besides, a central relay $R$ placed in the campus of University of Science and Technology of China (USTC) $(N31^{\circ}50^{'}8^{''}, E117^{\circ}15^{'}47^{''})$, is shared by all the users. The users' setup and the relay's BSM setup are shown in the inset.
In the user's side, we utilize an internally modulated signal laser, and modulate the decoy intensity according to vacuum+weak decoy scheme by an AM. Then we adopt a circulator, an asymmetrical Mach-Zehnder interferometer (AMZI), three amplitude modulators (AMs) and one phase modulator (PM) to encode qubits. The idle ports of circulators and beam splitters for the AMZI are exploited for phase synchronization and feedback, which are represented in the dash line. After attenuated by an electrical variable optical attenuator (EVOA), the signal laser pulses are sent via the deployed fiber to the relay comprised by an interference BS and two superconducting nanowire single-photon detectors (SNSPDs).
Before the BSM, we adopt a dense wavelength division multiplexor (DWDM) in each input of the BS to block the stray light in the fiber, and insert an electric polarization controller (EPC) and a polarization beam splitter (PBS) for polarization alignment.
}
\label{Fig:NetDeploySetup}
\end{figure*}

Here, we experimentally demonstrate a three-user, four-node MDIQKD network within the city of Hefei, China. The network deployment is shown in Fig.~\ref{Fig:NetDeploySetup}(a), which includes an untrusted relay, $R$, and three users, $U1$, $U2$, and $U3$, located in four different places. The deployed fiber lengths (channel losses) between the three users and the relay are 17 km (5.1 dB), 25 km (9.2 dB), and 30 km (8.1 dB), respectively. We employ an 8-by-4 mechanical optical switch to route the three users to the relay by auto-control command, with a low connection loss around 1 dB per channel.
The outputs of the switch are connected to a fiber beam splitter (BS) for  Bell state measurement (BSM). Within the star-type topology, any user pair ($U1-U2$, $U1-U3$, $U3-U2$) can get access to the BSM setup to run MDIQKD, which works as a quantum telephone exchange.

We emphasize that the MDIQKD network is not a straightforward upgrade of the previous point-to-point system\cite{Liu:MIQKDexp:2013,Tang:200kmMDIQKD:2014}.
There are significant new technical challenges in the network implementation.
The first one is reference frame calibration. Due to the polarization and phase fluctuation\cite{Tittel:MDIQKDFielfTest:2013,Liu:MIQKDexp:2013}, the reference frame needs to be calibrated timely. In the previous point-to-point experiments\cite{Liu:MIQKDexp:2013,Tang:200kmMDIQKD:2014}, we utilized an additional fiber link between the two users and one phase-stabilization laser (PSL) for this purpose. To scale up for networking applications, the demand of fiber links increases quadratically with the user number.
Besides, each new user needs an additional PSL with its wavelength locked to the signal laser.

\begin{figure*}[tbh]
\centering
\resizebox{12cm}{!}{\includegraphics{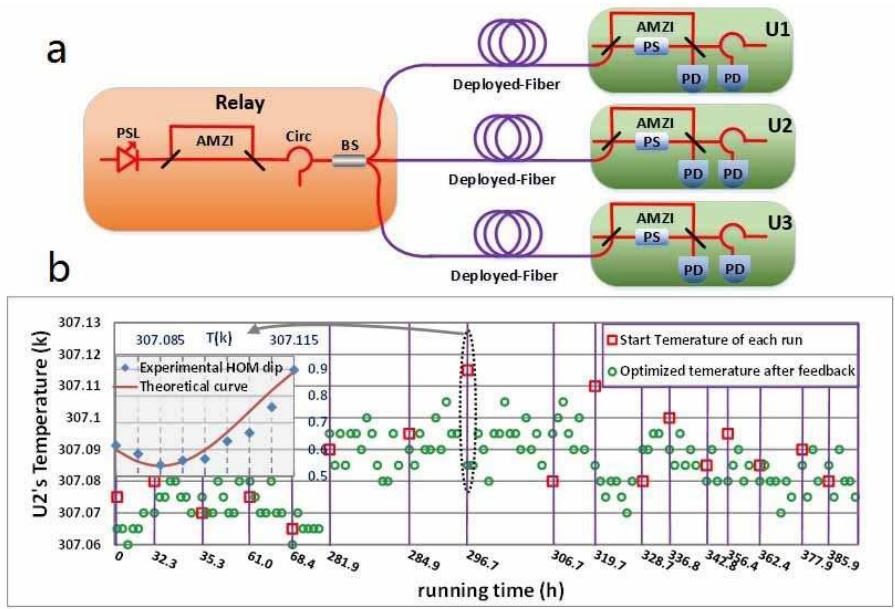}}
\caption{(a) The structure of phase-feedback scheme. A PSL placed in the relay passes through a reference AMZI with $6.5\ ns$ delay and a circulator. It is divided into three parts by BS and then sent to each user's AMZI connected by an additional fiber, respectively.
After the user's AMZI, two photon detectors (here we use commercial power meters) record the two output interference intensities, and then provide a feedback signal to compensate the phase shift of AMZI by the phase shifter inside each user's AMZI.
(b) The wavelength calibration result via measuring HOM interference. The different vertical zone represents different runs of MDIQKD. The X axis represents the system running time, and the Y axis represents the temperature for U2's laser in the case of user pair ($U3-U2$). Every green circle and red square point defines optimized temperature after a HOM dip measurement and the start temperature of each run. The inset shows the experimental HOM dip curve we measure by scanning the temperature.
}
\label{Fig:feedbackopt}
\end{figure*}

We use a scalable and efficient structure of phase stabilization suitable for phase calibration.
As shown in Fig.~\ref{Fig:feedbackopt}(a), a pulsed PSL is placed in the relay and passes through a reference asymmetrical Mach-Zehnder interferometer (AMZI) with a delay of $6.5\ ns$. It is divided by BS and then sent to the three users through three additional fibers, respectively.
After the user's AMZI, two commercial power meters record the two output interference intensities. Their measurement ratio provides a feedback signal to compensate the phase shift of AMZI by the phase shifter inside each user's AMZI.
In this arrangement, the required fiber resources increase only linearly with the user number, which is a significant reduction compared to that in point-to-point structure. In addition, only one PSL, placed in the relay and shared by all the users, is needed.

The other critical challenge is to maintain the indistinguishability of all the users' lasers. For a point-to-point system, a high-speed feedback system\cite{Tang:200kmMDIQKD:2014} was developed for this purpose where the two lasers are calibrated and locked via a feedback loop. Whereas, in the network, any two users will be switched upon their requests. Once switched, the two lasers must be calibrated immediately to guarantee their timing, spectrum and polarization mode are indistinguishable. Furthermore, the optical switch will change the lasers' arriving time and polarization when it is connected.

In our network, we randomly switch any two users to the relay BSM per two hours, i.e. we need to recalibrate all the lasers' modes per two hours. For the timing synchronization and polarization stabilization, we adopt the feedback system in our previous system\cite{Tang:200kmMDIQKD:2014} (see Methods). For the wavelength calibration, in the point-to-point system, we previously utilized an optical spectrum analyzer (OSA) to calibrate and feedback the wavelength. Here, we measure the Hong-Ou-Mandel (HOM) interference with the BSM setup and utilize the visibility of the HOM dip as the feedback signal. When the time and polarization are calibrated, the HOM visibility depends on the wavelength difference of the two lasers. And then we adjust the wavelength by tuning the temperature of the input signal lasers. As shown in the inset of Fig.~\ref{Fig:feedbackopt}(b),
by scanning the wavelength, we can observe a clear HOM dip. Using this method, we can take advantage of the existing BSM equipments, without the need of other instruments sush as the OSA. Furthermore, as the HOM dip can reflect the overall interference condition, it is also an efficient way to calibrate all the interference parameters, including the wavelength difference.

In our experiment, we optimize the system parameters\cite{Feihu:OptMDIQKD:2014}, including the decoy-state parameter and basis choice setting\cite{Wei2013DecoyBiased}, to increase the secure key rate. This optimization is based on the system parameters of the MDIQKD network with a high-quality transmitter system and a high-efficiency detection system. The MDIQKD transmitter at the user side and the BSM setup in the relay are illustrated in Fig.~\ref{Fig:NetDeploySetup}(b). Each user has the same configuration, adopting a signal laser internally modulated at 75 MHz, with a central wavelength of 1550.12 nm and a pulse width of 2.5 ns.
%
With decoy-state optimization, the laser pulse intensities of vacuum state, weak decoy state and signal state are 0, $\nu=0.1$ and $\mu=0.33$, respectively, and their probabilities are 16\%, 58\% and 26\%, respectively. Each user employs the time-bin phase-encoding scheme\cite{Ma2012AlterMDI}, in which only the raw data in the $Z$ basis are used for final key generation and those in the $X$ basis are used for phase error estimation. All the signal state is encoded in the $Z$ basis. For the weak decoy state, 63\% is encoded in the $X$ basis and the rest 37\% in the $Z$ basis. We utilize an AMZI, three AMs (AM2$\sim$AM4) and one PM to encode qubits. The modulators, AM2, AM3 and PM, are used to encode basis, and AM4 is used to normalize the two bases' average photon numbers. The signal laser pulses are attenuated to single-photon level by an EVOA.

After passing through the deployed fiber routed to the relay by an optical switch, the laser pulses are interfered in the BSM setup.
Before the BSM, we insert two DWDMs with 0.7 dB loss to block the background light. The BSM is then implemented with an interference BS and two SNSPDs\cite{sjchen:SNSPD:2015} operated at 2.1K with system detection efficiencies of 64\% and 66\% and the dark count rate of 100 Hz. The inner insertion loss of the BS is 1.4 dB. The partial BSM  post-selects $|\psi\rangle^-$ Bell state, when the two detectors in the two output arms of the BS have a coincidence detection at two alternative time bins.
We set an efficient time window of 1.7 ns to achieve a good interference.

With the BSM results announced by the untrusted relay, the two users run the basis sift by post selecting the raw data when they choose the same basis. Then the two users categorize the data according to vacuum, decoy and signal state labels and evaluate the gains and bit error rates in each case. All the data from the $Z$ basis, consisting of nine cases, will be used for  secure key extraction. The secure key rate formula is given by\cite{Lo:MIQKD:2012},
\begin{equation} \label{MIExp:Post:KeyrateMI}
\begin{aligned}
R &\ge \sum_{a,b\in\{0,\nu,\mu\}}Q_{11}^{ab}[1-H(e_{11}^{ab})] - Q^{ab} f H(E^{ab}),
\end{aligned}
\end{equation}
where $0,\nu,\mu$ denote the vacuum, decoy and signal states, respectively. $Q^{ab}$ and $E^{ab}$ are the overall gain and error rate when two users send states $a$ and $b$ with $a,b\in\{0,\nu,\mu,\}$. The gain and phase error rate of the single-photon components, $Q_{11}^{ab}$ and $e_{11}^{ab}$, can be estimated by the decoy-state method with a proper fluctuation analysis\cite{curty:finite:2014}. $H(e)=-e\log_{2}(e)-(1-e)\log_{2}(1-e)$ is the binary Shannon entropy function. The parameter $f$ is the error correction efficiency and we use $f=1.2$ for evaluation. The detailed key rate calculation is shown in Supplemental Material.

\begin{figure*}[tbh]
\centering
\resizebox{14cm}{!}{\includegraphics{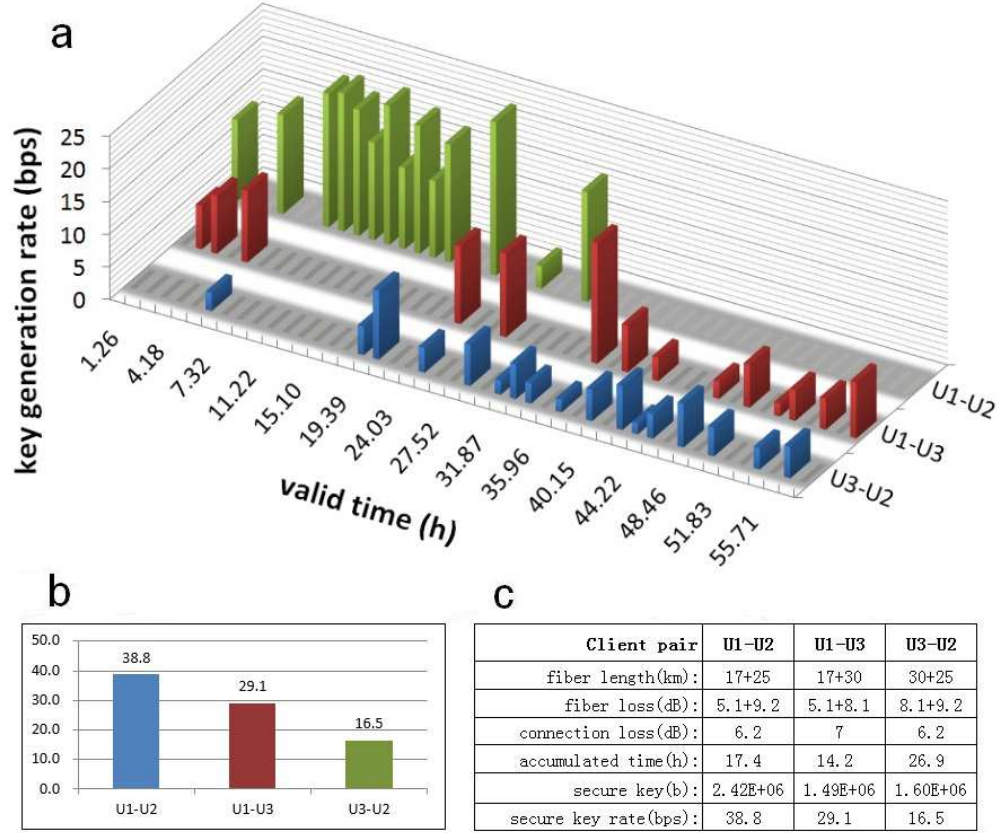}}
\caption{(a) the secure key rate array of each run with a valid time of $1.0\sim1.3$ hours for user pair ($U1-U2$), $0.8\sim1.2$ hours for ($U1-U3$) and $1.2\sim2.1$ hours for ($U3-U2$).
(b) the overall key rate (unit: bps) with accumulated data of each user pair in 17.4 hours ($U1-U2$), 14.2 hours ($U1-U3$) and 26.9 hours ($U3-U2$).
(c) the system parameters, including the loss, the accumulation time and secure key rate obtained.
}
\label{Fig:KeyrateResult}
\end{figure*}

We run the post-processing for each valid data session of 1 to 2 hours between different user pairs. In the analysis, we fix the failure probability to be $10^{-10}$. The secure key rates between different pairs, $U1-U2$, $U1-U3$, and $U3-U2$, in different runs are shown in Fig.~\ref{Fig:KeyrateResult}(a). They are averagely 17.1 bps in 1 hour ($U1-U2$), 6.4 bps in 1 hour ($U1-U3$) and 4.2 bps in 1.5 hours ($U3-U2$). Furthermore, we analysis the secure key rate by accumulating all the valid data, and have extracted 38.8 bps in 17.4 hours ($U1-U2$), 29.1 bps in 14.2 hours ($U1-U3$) and 16.5 bps in 26.9 hours ($U3-U2$), as shown in Fig.~\ref{Fig:KeyrateResult}(b).
We remark that the results we have achieved are at least ten times higher than the previous state-of-the-art field test.

In conclusion, we have demonstrated the first MDIQKD network which is secure against untrustful replays and all detection attacks, and also resource efficient in real world implementation. The multi-user HOM interference technology developed in the experiment can find applications in multi-party entanglement swapping based quantum communication\cite{bose:multiparticle:1998} and a quantum computing cloud. In quantum computing cloud, the users only need to prepare quantum states and share the expensive quantum computing devices. Furthermore, with the decoy state source, such topological setup can also be extended to the blinding quantum computing\cite{PhysRevLett.108.200502}, where the computing station can be untrusted.

\textbf{Methods}

\textbf{Fully-automated feedback system.} The time calibration system mainly includes synchronization laser (SynL, 1570 nm) pulses to synchronize the whole system, and programmable delay chip to adjust the time delay of SynL pulses.
The polarization stabilization system in each arm of the interference BS in the BSM handles the polarization misalignment of the laser pulse connected to BSM by the optical switch. It mainly includes an EPC, a PBS,
a SNSPD
in the reflection port of this PBS, and a polarization-maintained interference BS in the transmission port.

For the wavelength calibration, we implement the HOM interference and calculate the coincidence value of HOM dip, rather than measure the wavelength difference directly by an OSA.
The coincidence value of HOM dip is calculated by $(N_{c}N_{tot})/(N_{1}N_{2})$, where $N_{c}$, $N_{tot}$, $N_{1}$ and $N_{2}$ represent the coincidence count of two BSM detectors, the total pulse count sent by the laser source, the detection count of BSM detector 1 and 2, respectively, over a certain run time and within a certain time window.
By scanning the temperature which increases linearly with the wavelength of our laser ($0.8\ pm$ per $0.01$ degree centigrade), we can obtain the optimized wavelength.
To implement HOM dip measurement, the two users send their strong laser pulses without any decoy or qubit modulation. Besides, it's preferable for the intensity of the laser pulses arriving at the BSM setup to be close to each other. This is because the coincidence value represents the indispensability considering all the modes. To obtain the wavelength difference more precisely, the coincidence value should be less influenced by the other aspects.
To fulfill this purpose, the EVOA in the user's side is utilized to adjust the output intensity in this wavelength calibration procedure.

For the phase stabilization, we adopt a pulsed PSL (1550.12 nm) with 2.5 ns pulse width placed in the relay. It passes through an AMZI with 6.5 ns time difference of two paths. Then, a BS divides the pulsed laser into three parts, with each output port connecting to one user.
Combined with the synchronization laser pulses by WDM,
The PSL pulses are transmitted through an additional fiber links and received by the corresponding user.
Separated by another WDM, the PSL pulses passes through the user's AMZI, and are monitored by two commercial power meters in the two interference output. The intensity ratio provides a feedback signal to calibrate two AMZIs' phase reference frame by the phase shifter inside each user's AMZI.
This arrangement has two advantages in both the scalable structure and the low commercial cost. Firstly, this new structure is preferred for extension, with only one PSL placed in the relay and shared by all the users.
In contrast, the structure of phase stabilization system suitable for point-to-point implementation\cite{Tang:200kmMDIQKD:2014} needs one more feedback laser source when one user joins in the network, and all the feedback laser wavelengths should be locked to the signal laser, which increases the technical complexity.
Secondly, we adopt the commercial power meter rather than the gated single-photon detector. On one hand, the power meter is much cheaper, especially preferred by users. On the other hand, the single-photon detector requires a gate signal with time calibrated according to fiber length shift. However, the power meter requires no extra controlling signal and thence no calibration to it.

\textbf{\subsection*{Acknowledgments}}
The authors would like to thank J. Fan, X.~Xie, Z.~Cao and M.~Jiang for enlightening discussions, as well as D.~Yang and W.~Sun for experimental assistance. This work has been supported by the National Fundamental Research Program (under Grant No. 2011CB921300 and 2013CB336800), the National Natural Science Foundation of China, the Chinese Academy of Science, and the Shandong Institute of Quantum Science \& Technology Co., Ltd.






\newpage

\end{document}